\documentstyle[11pt]{article}
\input epsf

\begin{document}
\def\beq{\begin{equation}}
\def\eeq{\end{equation}}
\def\tt{\tilde{T}}
\def\ta{\tilde{A}}

%\rightline{USC -???}
\rightline{  }
\vskip30pt

\begin{center}
{\Large\bf Correlations between $<p_T>$ and jet multiplicities\\
\vskip15 pt 
from the BFKL chain}	
\vspace {1 cm}

M.A.Braun\footnote{On leave of absence from St.Petersburg 
State University (Russia)},
C.Merino and G.Rodr\'\i guez
\vspace{0.5 cm}

{\it Department of Particle Physics, University of Santiago de Compostela,\\
15706 Santiago de Compostela, Galiza, Spain}
\end{center}

\vskip30pt
\begin{abstract}
Strong correlations between the number of emitted jets and their average 
transverse momentum are found for the events
resulting from the exchange of a single BFKL pomeron.
\end{abstract}

%\section{Introduction}
As  well-known, strong correlations are observed experimentally 
between the average 
$p_T$ and multiplicities of particles produced in high-energy hadronic
collisions [1]. Average $p_T$ grows with multiplicity. This fact can be
interpreted as a consequence of multiple hard collisions, which result in
a larger number of particles produced and a broadening of the $p_T$ spectrum
[2]. An alternative interpretation can be given in terms of colour 
strings which
are stretched between the colliding hadrons during the collision. 
Here larger
multiplicities correspond to a larger number of strings, which again 
leads to a
broadening of the $p_T$ spectrum either because of the accumulation of more
transverse momentum from the parent partons or because of the interaction
between strings [3]. In both cases it is tacitly assumed that with only 
one hard
collision or, alternatively, with only one pair of non-interacting strings
there are no correlations between $<p_T>$ and multiplicity. Theoretically
this assumption can only be tested within the BFKL dynamics, which presents a
detailed description of particle (actually jet) production at high energies
under certain simplifying assumptions (a fixed small coupling constant).
The present calculation is aimed to see whether there exist correlations between
$<p_T>$ and the number of produced jets in the hard pomeron 
described by the BFKL
chain of interacting reggeized gluons.

Our study is closely related to the paper by J.Kwiecinski, C.A.M.Lewis and
A.D.Martin,
who calculated the exclusive probabilities to observe a given number of jets
from the exchanged hard pomeron [4]. We shall extensively use both their 
method of calculations and some of their parametrizations.

The results we present in this letter demonstrate that, in fact, 
at realistic energies in the BFKL 
chain there 
are strong correlations  between the number $n$ of produced jets and
the average $<p_T>_n$ for these events. 
We find that  $<p_T>_n$ grows roughly linearly with n
with a slope independent of $Q^2$ in deep inelastic scattering
(DIS) and of the same value for purely hadronic collisions. 
The slope diminishes with
rapidity $y$.
So in the limit $y\rightarrow\infty$ the correlations are expected to 
disappear.
However, at $y=15$ ($x\sim 10^{-7}$) the value of the slope is 
still $\sim 0.8$ GeV/c.
These results leave open the question of  correlations between $<p_T>_n$ 
and $n$ in the soft pomeron and in the colour string at the currently available energies.

%\section{Formalism}
We start by recalling the main points of the formalism employed in [4] to 
calculate the exclusive probabilities for the production of a given number of
jets. To facilitate comparison with [4] we shall borrow their notations.
Let the amputated BFKL amplitude be $f(y,k)$, when 
$y$ is the rapidity and $k$ is
the 2-dimensional transverse momentum of the virtual (reggeized) gluon.
Function $f(y,k)/k^2$ is interpreted as an unintegrated gluon distribution,
related to the conventional gluon distribution by
\beq
xG(x,Q^2)=\int^{Q^2}\frac{dk^2}{k^2}f(\ln\frac{1}{x},k).
\eeq
The BFKL equation for $f$ may be written in the form
\beq
f(y,k)=f^{(0)}(y,k)+\bar{\alpha}_s\int_0^ydy_1\int\frac{d^2k_1}{\pi q^2}
\Big(\frac{k^2}{k_1^2}f(y_1,k_1)-f(y,k)\theta (k^2-q^2)\Big).
\eeq
Here $\bar{\alpha}_s=3\alpha_s/\pi$, $q=k-k_1$ is the transverse 
momentum of the emitted (real) gluon and it was assumed in [4] that the 
driving
term $f^{(0)}$ (the impact factor of the target) may also depend on rapidity.
To suppress the physically unknown infrared domain and make the equation
numerically tractable, the integration over $k_1$ was constrained in [4] 
to the interval
\beq
Q_0^2<k_1^2<Q_f^2,
\eeq
with $Q_0=1 $ GeV/c and $Q_f=100$ GeV/c. We shall also impose this constraint,
implicit in the following equations.

Defining as an observable jet a real gluon with $q^2\geq \mu^2$, one
splits the integration over momenta in (2) into two parts by introducing
\beq
\theta(q^2-\mu^2)+\theta(\mu^2-q^2)=1
\eeq inside the integral.
The whole integration kernel is thus split into two parts: a resolved one,
$K_R$, corresponding to emitted gluons with $q^2>\mu^2$, and an unresolved 
one, $K_{UV}$,
which combines emission of gluons with $q^2<\mu^2$ and the subtraction term in
(2). Explicitly, the action in the momentum space of the two kernels is 
described by \beq
\Big(K_Rf\Big)(k)=\bar{\alpha_s}k^2\int\frac{d^2k_1}{\pi q^2 k_1^2}
\theta(q^2-\mu^2)f(k_1),
\eeq
\beq
\Big(K_{UV}f\Big)(k)=\bar{\alpha_s}k^2\int\frac{d^2k_1}{\pi q^2 k_1^2}
\Big(\theta(\mu^2-q^2)f(k_1)-\frac{k_1^2}{k^2}\theta(k^2-q^2) f(k)\Big).
\eeq

Exclusive probabilities to produce $n$ jets are obviously obtained by 
introducing $n$ operators $K_R$ between the Green functions of the BFKL 
equations with kernel $K_{UV}$. If one presents the full gluon distribution
$f$ as a sum of contributions $f_n$ from the production of $n$ jets
\beq
f(y)=\sum_{n=0}f_n(y),
\eeq then one gets  a recursive relation
\beq
f_n(y)=\int_0^ydy_1K(y-y_1)f_{n-1}(y_1),
\eeq
where $K(y)$ is an $y$-dependent operator in the transverse momentum space
\beq
K(y)=e^{yK_{UV}}K_R.
\eeq
Eq. (8) allows to successively calculate the relative probabilities to 
produce $n=0,1,2,...$ jets starting from the no-jet contribution
determined by
\beq
f_0(y)=e^{yK_{UV}}f^{(0)}(0)+
\int_0^ydy_1e^{(y-y_1)K_{UV}}\frac{df^{(0)}(y_1)}{dy_1}.
\eeq
In [4] the driving term was chosen to vanish at $y=0$:
\beq
f^{(0)}(y,k)=A(1-e^{-y})^5e^{-k^2/Q_0^2}
\eeq
(its normalization factor is irrelevant for our purpose).

Distributions $f_n(y,k)$ themselves are not observable quantities. Physical
probabilities are obtained by convoluting $f_n$ with the gluon distribution
in the projectile (the projectile impact factor). For the perturbative 
QCD to be applicable, a reasonable choice is to take the virtual 
photon  as a projectile, as done in [4]. Having in mind that the
BFKL picture may only be applied to low values of $x$, in our 
calculations we used a simplified expression for the virtual photon
impact factor, independent of rapidity, which can be found in 
[5]. To have some qualitative idea of the situation in purely
hadronic collisions, we have also made our calculations for  
a hadronic projectile with an unperturbative impact factor.
For collisions of two identical hadrons 
it should be identical to the target impact factor
which appears as an $y$-independent driving term
$f^{(0)}(k)$ in (2). 

In both cases the exclusive probabilities to 
observe $n$ jets are given by
\beq
P_n(y)=\frac{\int (d^2k/k^4)h(k)f_n(y,k)}
{\int (d^2k/k^4)h(k)f(y,k)},
\eeq
where $h(k)$ is the impact factor of the projectile.
Both impact factors,  $f^{(0)}(k)$ of the target and $h(k)$
of the projectile, should vanish as 
$k\rightarrow 0$. This condition is satisfied by the virtual photon
impact factor of [5]. As to the hadronic impact factor
$f^{(0)}(k)$, we have chosen it in close similarity with (11):
\beq
f^{(0)}(k)=k^2e^{-k^2/Q_0^2}.
\eeq
As with Eq. (11), the overall normalization  is irrelevant.

We are interested in the average values of $<q>_n$ in the observed jets,
provided their number $n$ is fixed. At this point one has to remember that the momentum 
$k$ which serves as an argument of $f(y,k)$ refers to the virtual gluon, 
and not to the emitted one, whose momentum $q$ is hidden inside the kernel
$K_R$. Therefore, to find an average of any quantity $\phi(q)$ depending 
on the emitted real jet momentum, one has to introduce the function $\phi(q)$
into the integral (5), thus changing the kernel $K_R$ to the kernel $K_{av}$ defined by
\beq
\Big(K_{av}f\Big)(k)=\bar{\alpha_s}k^2\int\frac{d^2k_1}{\pi q^2 k_1^2}
\theta(q^2-\mu^2)\phi(q)f(k_1).
\eeq
With $n$ jets, one has to substitute
one of the $n$ operators $K_R$ which generate the jets by $K_{av}$, take a sum of all 
such substitutions, and divide by $n$. One has further to integrate over 
all momenta of the virtual gluon $k$ multiplied by the projectile impact factor,
and normalize the result to the total probability to have $n$ jets. 

This recipe can be formalized in the following way. Introduce a 
generalized operator in the virtual gluon momentum space
\beq
K_1(y)=e^{yK_{UV}}[K_R+K_{av}].
\eeq
Let the function $F(y,k)$ obey the equation
\beq
F(y)=f_0(y)+\int_0^ydy_1K_1(y-y_1)F(y_1).
\eeq
One can split the function $F$ into a sum of contributions $F_{nm}$
corresponding to the action of  $n$ operators $K_1$, out of which
$m=0,1,...n$, are operators $K_{av}$:
\beq
F(y)=\sum_{n=0}\sum_{m=o}^n F_{nm}(y).
\eeq
Evidently $F_{n0}=f_n$. We are interested in the contribution
$F_{n1}\equiv g_n$ which contains a single operator $K_{av}$. The average 
value of interest
is determined by
\beq
<\phi(q)>_n=\frac{1}{n}\frac{\int (dk^2/k^4)h(k)g_n(y,k)}
{\int (dk^2/k^4)h(k)f_n(y,k)}.
\eeq

In analogy with (8), one easily sets up a recursion relation for $g_n$:
\beq
g_n=\int_0^ydy_1K(y-y_1)g_{n-1}(y_1)+\int_0^ydy_1
e^{(y-y_1)K_{UV}}K_{av}f_{n-1}(y_1),
\eeq
with the initial condition $g_0(y)=0$. Together with (8), this relation
allows to calculate the function $g_n$ for $n=1,2,...$, and then to use (18) to
find the desired averages.

The concrete choice of $\phi(q)$ is restricted by the condition of
convergence at large $q$: $\phi(q)<q^2$, as $q\rightarrow\infty$.
For this reason we take $\phi(q)=q$, so that
\beq
\Big(K_{av}f\Big)(k)=\bar{\alpha_s}k^2\int\frac{d^2k_1}{\pi k_1^2q}
\theta(q^2-\mu^2)f(k_1).
\eeq

%\section{Numerical results and discussion}
We defined our jets by taking $\mu=2$ GeV/c. As for the cutoffs,
we used (3). To see the influence of the cutoffs we also repeated
our calculations with $Q_0=0.5$ GeV/c and $Q_f=1000$ GeV/c. 
The results slightly 
change in their 
absolute values (by no more than $1$-$6$\%) but both the $n$ and $y$ 
dependences remain the same.

We have calculated the functions $f_n$ and $g_n$  from Eqs. (8) and (19)  
up to $n=5$ and $y=15$. Following [4]
we have used the expansion in $N$ Chebyshev polynomials to discretize the
kernels in a simple way. 
The results we present have been obtained
with $N=80$, although, as pointed out in [4], already  $N=20$ gives a
reasonable approximation.  

In Figs. 1-3 we present the averages $<q>_n$ 
for $n=1$-$5$ and $x=e^{-y}=3.10^{-7}$-$0.1$, 
for the $\gamma^*$-hadron collisions at $Q^2=10$, 100 and 
1000 (GeV/c)$^2$.
In Fig 4 we show these averages for the collisions of two identical
hadrons with the impact factor (13).
As one observes, in all cases $<q>_n$ strongly grows with $n$ at all 
rapidities. The growth is approximately linear 
\beq
 <p_T>_n\simeq a(y,Q^2)+b(y)\,n,
\eeq
where both $a$ and $b$ depend on rapidity $y$, but, at fixed $y$, $b$ is
universal in the sense that it does not depend on $Q^2$ in DIS and
has the same value for pure hadronic collisions.
The slope $b(y)$ falls with $y$: it is equal 1.25 GeV/c at $y=7.5$, and 
0.8 GeV/c at $y=15$, so that one may expect that at ultrahigh 
energies $<q>_n$ will become independent of $n$.

As an interesting byproduct of our study we find that the averages $<q>_n$
go down with rapidity for all $n\geq 2$. This is quite unexpected, since,
as well-known, in the BFKL approach an overall average $<q>$ rapidly 
grows with $y$ ([6] and Eq. (25) below).
It seems that this growth is totally explained by the growth of
the average number of jets $<n>$. 

Passing to discussion, we first point out that 
it is an open question in which kinematical conditions and to what 
degree 
the BFKL pomeron may describe realistic hadronic processes. Emissions of 
high-$p_T$ jets in $\gamma^*$-hadron collisions seem to be a suitable 
place to
see the BFKL signatures. Our results show that in such emissions strong
positive correlations are predicted between $<p_T>$ and the number of jets,
already for a single pomeron exchange. This indicates that in fact such
correlations do not require 
multiple rescatterings nor pomeronic interactions, but they are
already present in the basic mechanism of jet production. 
Obviously, this conclusion cannot be directly applied to particle 
production in the soft region, and so the empirically invoked absence of such 
correlations for particle production from the colour string does not really
contradict our results. It can only be tested 
in its own framework by
confronting colour string predictions with the experimental data.

Our result (21) has been obtained for relatively small number of jets
and at energies corresponding to $y\leq 15$. If extrapolated to all $n$ 
and energies it would lead to a relation between the overall averages
\beq
<p_T>\propto <n>.
\eeq
However, it is well-known that this relation does not hold
in the BFKL model
at asymptotic energies, when $<p_T>$ grows much faster than $<n>$.
This more or less known fact can be demonstrated by a simple calculation.
Indeed 
the inclusive cross-section $I(y,y_1,q)$
to produce a jet at rapidity $y_1$ and with a transverse momentum $q$,
in a collison with an overall rapidity $y>>1$ is given by [7]
\beq
I(y,y_1,q)=\frac{\bar{\alpha}_s}{4\pi}\sigma(y)\frac{1}{q^2}
\Big(1-\Phi(z)\Big),
\eeq
where
\beq
z=b\ln q,\ \ b=\sqrt{\frac{y}{ay_1(y-y_1)}},\ \ a=14\bar{\alpha}_s\zeta(3).
\eeq
Here $\Phi$ is the error function and $\sigma(y)$ is the total cross-section.
Since $b\sim 1/\sqrt{y}<<1$, the term with $\Phi$ is actually important
only at large $\ln q$ when it cuts the distribution in $q$ at
$\ln  q\sim \sqrt{y}$. For this reason the scale of $q$ is unimportant, so
that one may safely fix it by setting 
$\mu=1$.                                                  
The average value of any positive power $\beta$ of the transverse momentum
is easily found to be
\beq
<p_T^{\beta}>=\frac{\int_0^ydy_1\int d^2qq^{\beta}I(y,y_1,q)}
{\int_0^ydy_1\int d^2q I(y,y_1,q)}=
\frac{1}{\lambda^2}e^{\lambda^2}\Phi(\lambda),\ \
\lambda=(1/2)\beta\sqrt{ay}.
\eeq
Evidently $<p_T^{\beta}>$ grows exponentially with $y$
for any $\beta>0$.                                                                                                                                                   

Thus  our results cannot be valid for all $n$ and 
energies  and refer precisely to the values of $n$ and energies
for which the calculations were done.  
It is interesting to note that relations similar to (22), with $<p_T>$
substituted by $<p_T^2>$, were earlier obtained
from gluon saturation [8], and in the percolation approach [9]. 

As mentioned, an unexpected result obtained in our calculation is 
that $<p_T>_n$ at fixed $n\geq 2$ fall with energy. 
As  seen in Figs. 1-4 this fall is not dramatic at energies at which we
can expect the BFKL pomeron to be relevant ($y>10$). Still, it is quite
appreciable, especially in view of the naive belief that the average
transverse momentum should grow with energy. This 
prediction can easily be tested experimentally as a possible signature of the
BFKL pomeron.
\vspace{0.5 cm}

It is a pleasure to thank C. Pajares for interesting discussions and for
pointing out to us some relevant references. 
We also thank J. Castro and G. Parente for their useful comments.
This work was supported by the Secretar\'\i a de Estado de Educaci\'on y 
Universidades of Spain and by the RFFI grant
01-02-17137 (Russia).
\vspace{0.5 cm}

{\bf References}

\noindent
1. UA1 Collaboration, C.Ciapetti in {\it The Quark Structure of 
Matter}, edited by 
M.Jacob and K.Winter (1986) 455;
F.Ceradini, Proc. Int. Europhys. Conference on High-Energy Physics,
Bari, edited by L.Nitti and G.Preparata (1985), and references therein.

\noindent
%2. {\it Hard scattering explanations}
2. A.Capella, J.Tran Thanh Van and J.Kwiecinski, Phys. Rev. Lett.
{\bf 58} (1987) 2015.

\noindent
%{\it DPM + ministrings}
3. A.Capella and A.Krzywicki, Phys. Rev. {\bf D 29} (1984) 1007;
P.Aurenche, F.W.Bopp and J.Ranft, Phys. Lett. {\bf B 147} (1984) 212;
C.Merino, C.Pajares and J.Ranft, Phys. Lett. {\bf B 276} (1992) 168;
M.A.Braun and C. Pajares, Phys. Rev. Lett. {\bf 85} (2001) 4864. 

\noindent
%{\it And string interactions ?}
4. J.Kwiecinski, C.A.M.Lewis and A.D.Martin, Phys. Rev {\bf D 54} 
(1996) 6664.

\noindent
5. N.N.Nikolaev and B.G.Zakharov, Z.Phys. {\bf C 49} (1991) 607.

\noindent
6. L.V.Gribov, E.M.Levin and M.G.Ryskin, Phys. Rep. {\bf 100} (1983) 1.

\noindent
7. M.A.Braun and D.Treleani, Eur. Phys. J. {\bf C 4} (1998) 685

\noindent
8. J.Schaffner-Bielich, D.Kharzeev, L. McLerran and R.Venugopalan,
nucl-th/0108048;
L.McLerran and J.Schaffner-Bielich, Phys. Lett. 
{\bf B 514} (2001) 29.

\noindent
9. M.A.Braun, F.del Moral and C.Pajares, hep-ph/0105263.
\vspace{0.5 cm}

{\bf Figure captions}

\noindent
Figure 1.  Average $<p_T>_n$  for a  fixed number 
$n$ of 
jets produced in $\gamma*$-hadron collisions, as a function of 
$x$  at $Q^2=10$ (GeV/c)$^2$. Curves from bottom to top
correspond to $n=1,2,...5$.
\noindent
Figure 2.  Average $<p_T>_n$  for a  fixed number
$n$ of
jets produced in $\gamma*$-hadron collisions, as a function of
$x$  at $Q^2=100$ (GeV/c)$^2$. Curves from bottom to top
correspond to $n=1,2,...5$.                                                     
\noindent
Figure 3.  Average $<p_T>_n$  for a  fixed number
$n$ of
jets produced in $\gamma*$-hadron collisions, as a function of
$x$  at $Q^2=1000$ (GeV/c)$^2$. Curves from bottom to top
correspond to $n=1,2,...5$.                                                     
\noindent
Figure 4. Average $<p_T>_n$  for a  fixed number
$n$ of jets produced in hadronic collisions, as a function 
of   $y$ . Curves from bottom to top 
correspond to $n=1,2,...5$.

%%%%%%%%%%%%%%%%%%%%%%%%%%%%%%%%%%%%%%%%%%%%%%%%%%%%%%%%%%%%%%%%%%%%%%%%
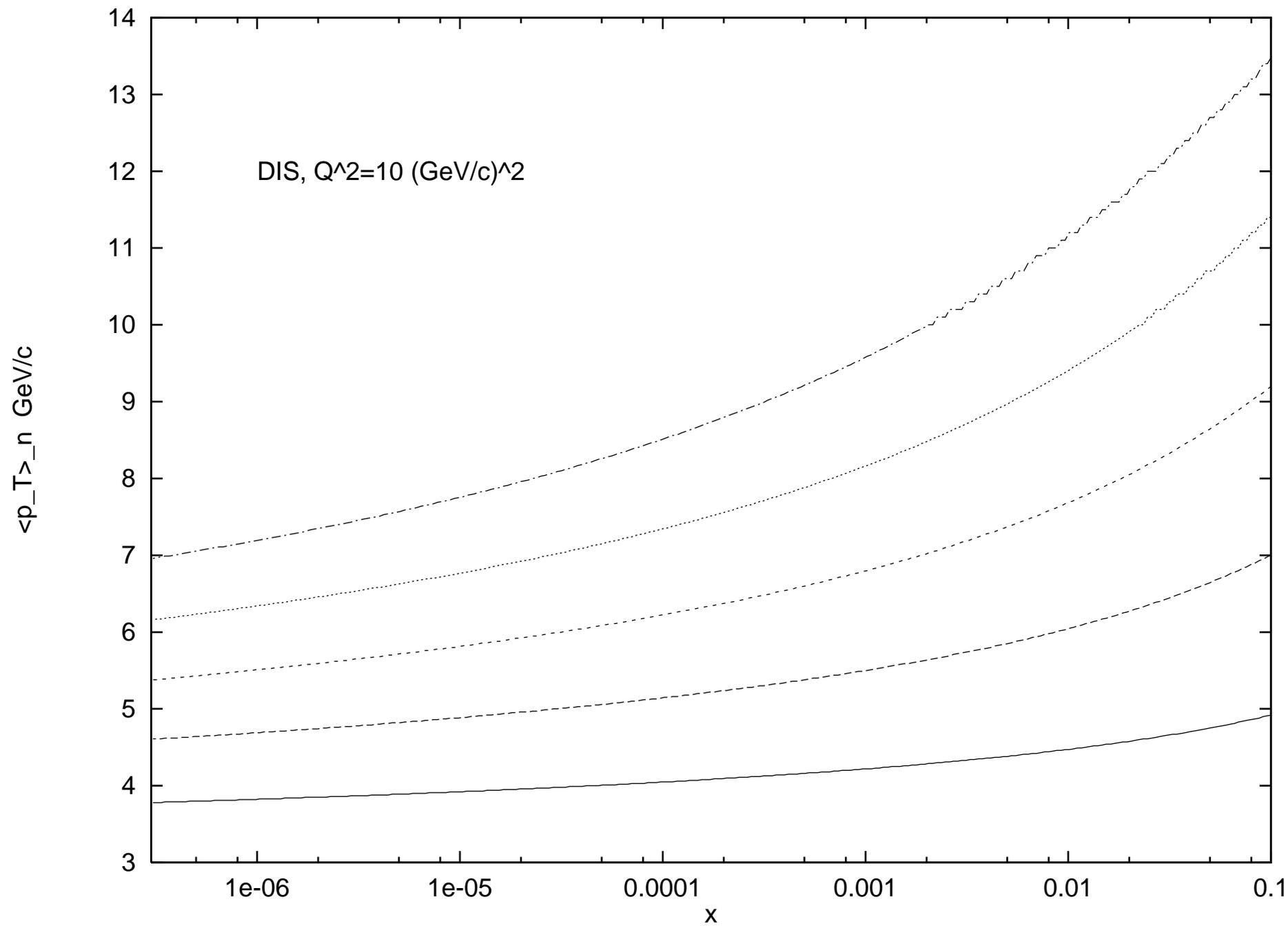
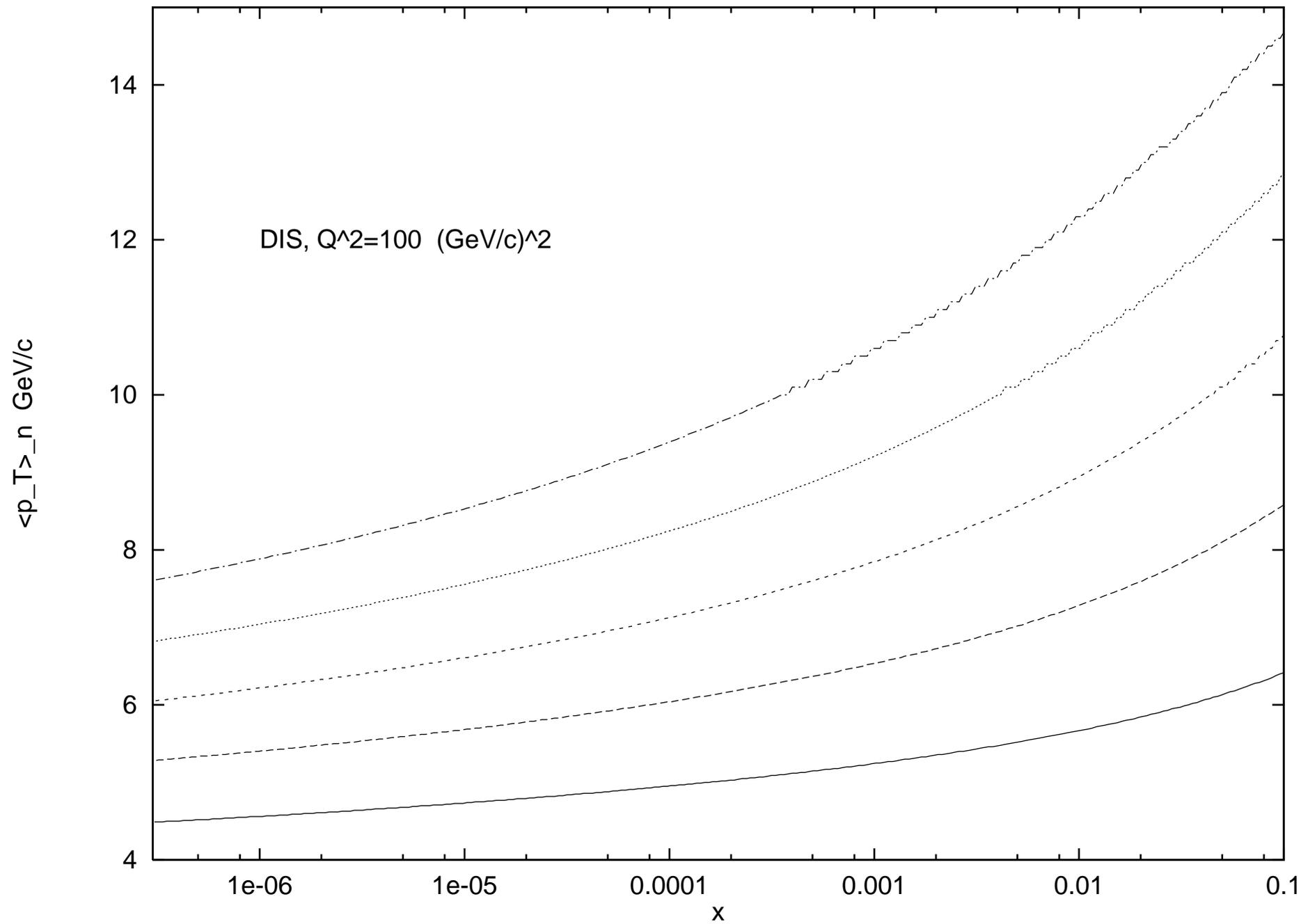
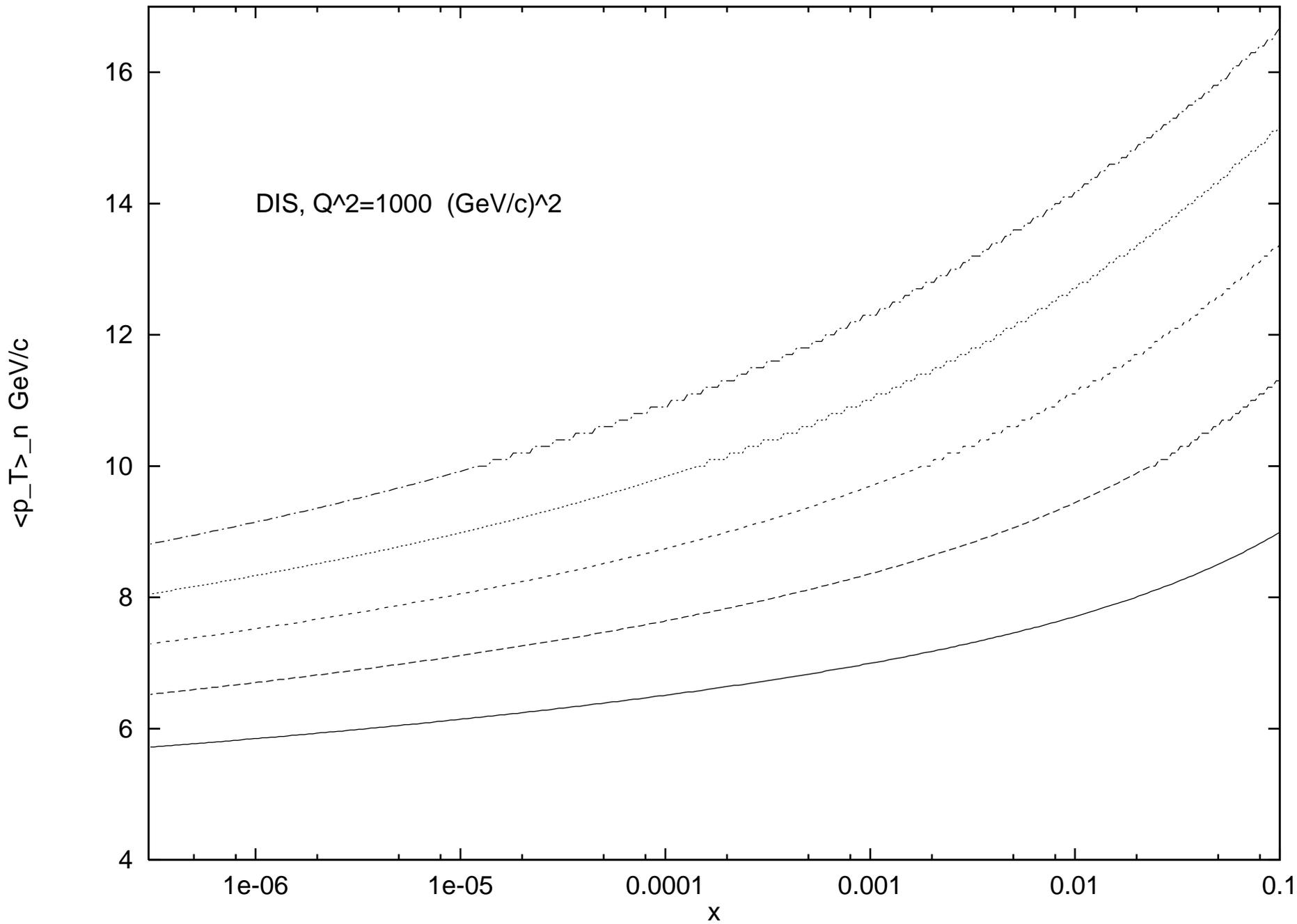
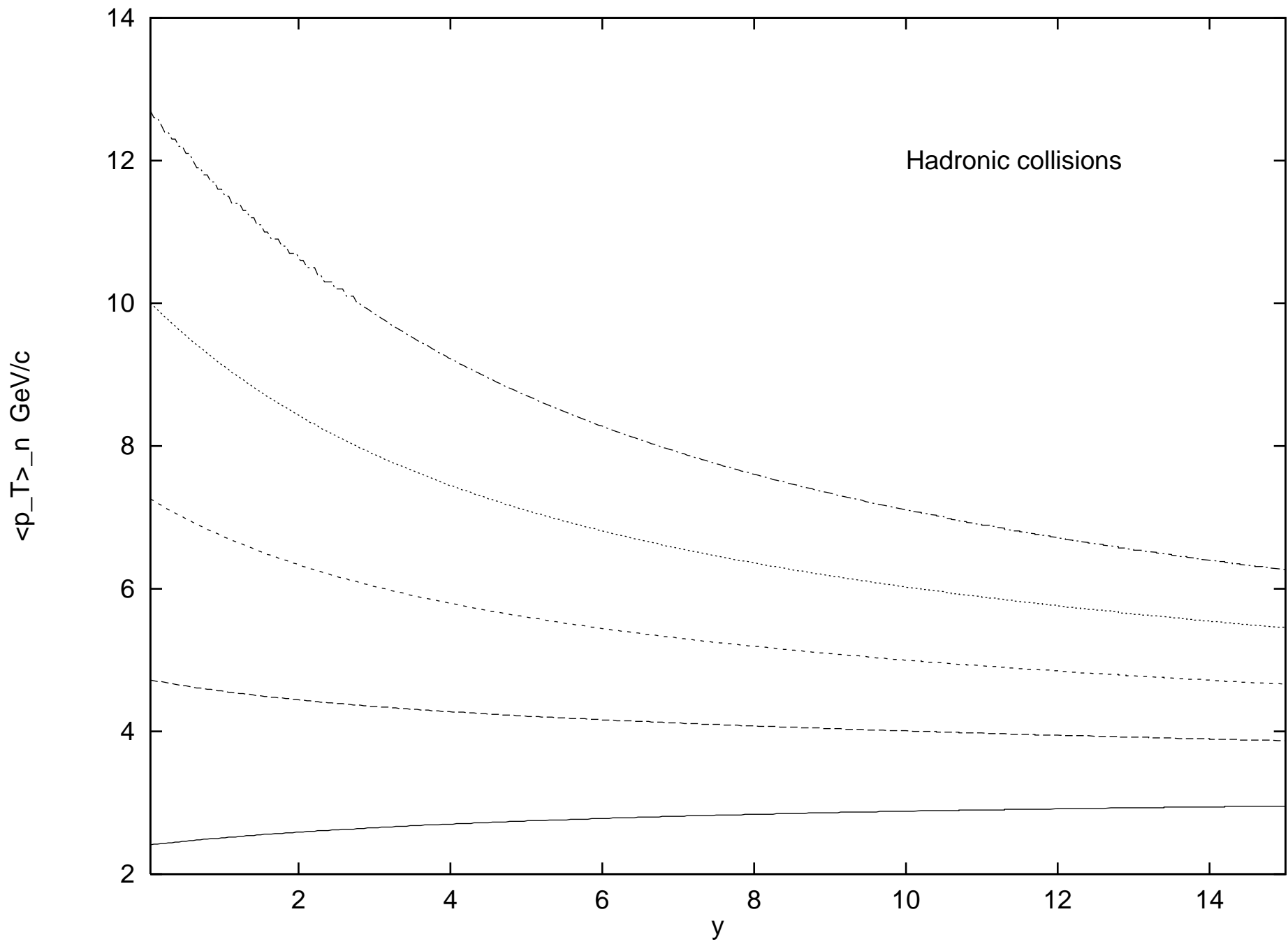
\end{document}